\documentclass[conference]{IEEEtran}
\IEEEoverridecommandlockouts
\usepackage{cite}
\usepackage{amsmath,amssymb,amsfonts}
\usepackage{algorithmic}
\usepackage{graphicx}
\usepackage{textcomp}
\usepackage{xcolor}
\usepackage{enumitem}
\usepackage{booktabs}
\usepackage{multirow}
\usepackage{bm}
\usepackage{url}
\usepackage{balance}
\def\BibTeX{{\rm B\kern-.05em{\sc i\kern-.025em b}\kern-.08em
    T\kern-.1667em\lower.7ex\hbox{E}\kern-.125emX}}
\begin{document}

\title{Optimization of Predictive Maintenance Schedules under Uncertainty:
A Scenario-Based Theoretical Framework
\thanks{Funded by the European Union. Views and opinion expressed are however those of the author(s) only and do not necessarily reflect those of the European Union or Europe’s Rail Joint Undertaking. Neither the European Union nor the granting authority can be held responsible for them. The project FP3-IAM4Rail is supported by the Europe's Rail Joint Undertaking and its members.}
}

\author{\IEEEauthorblockN{Jerzy Baranowski, Waldemar Bauer\\}
\IEEEauthorblockA{\textit{Department of Automatic Control and Robotics}\\
\textit{AGH University of Science and Technology}\\
 Krakow, Poland\\\
\{jb, bauer\}@agh.edu.pl}
}

\maketitle

\begin{abstract}
This paper proposes a scenario-based framework for predictive maintenance scheduling under uncertainty in a finite planning horizon. The considered setting involves multiple assets for which maintenance decisions are informed by three heterogeneous sources of information: calendar-based overhaul intervals, usage-based limits driven by uncertain future operating cycles, and condition-monitoring outputs represented through remaining useful life (RUL) estimates with uncertainty. While these elements have been studied extensively in the maintenance literature, they are often treated separately or only partially integrated. In contrast, the proposed formulation evaluates complete maintenance schedules under simulated future scenarios and compares them using expected-cost and tail-risk criteria. The contribution is primarily conceptual and methodological: we define a unified finite-horizon decision framework that combines calendar-, usage-, and prognostics-based information within a common scheduling problem. A small synthetic computational example is used as a proof of concept. The results show that integrated scenario-based policies can substantially outperform simpler single-trigger rules, while the difference between risk-neutral and risk-aware integrated policies remains modest under the present calibration.

\end{abstract}


% Cell 8: replacement for Section I (Introduction)

% Cell 18: replacement for Section I (Introduction)

\section{Introduction}

Maintenance planning in engineered systems is increasingly expected to use predictive information rather than rely solely on fixed service intervals. In practice, however, maintenance decisions are rarely driven by a single signal. Asset managers often need to reconcile several heterogeneous sources of information, such as elapsed time since the last intervention, accumulated or forecasted usage, and condition-monitoring indicators related to degradation. This creates a scheduling problem that is richer than a standard threshold-based maintenance rule and naturally raises questions about how these signals should be combined within a unified decision framework.

The present paper is motivated by precisely such a setting. We consider a finite-horizon maintenance scheduling problem for multiple assets, where each asset is described by three maintenance-relevant quantities: a calendar-based resource, a usage-based resource expressed in operating cycles, and a condition-monitoring signal summarized by an estimate of the remaining useful life (RUL) with uncertainty. The decision problem is to determine a complete maintenance schedule over the planning horizon rather than to issue a local next-step recommendation for a single asset.

This setting is practically relevant because these three types of information often coexist in real maintenance environments. Some interventions remain anchored in prescribed overhaul intervals, some are driven by cumulative usage, and some are supported by diagnostic or prognostic information. At the same time, these signals differ not only in meaning but also in uncertainty structure. Calendar information is usually deterministic, usage depends on future operating conditions, and condition-based information is inherently probabilistic. As a result, the maintenance planner faces an uncertainty-aware scheduling problem rather than a deterministic inspection or replacement problem.

The existing literature provides substantial building blocks for such a formulation, including age-based maintenance, condition-based maintenance, prognostics-driven decision-making, multi-component maintenance optimization, and risk-aware maintenance criteria. However, these strands remain only partially integrated. Many studies focus on one dominant trigger, a limited hybridization of two information sources, or expected-value optimization without explicit schedule evaluation under a distribution of future outcomes. This leaves a clear methodological gap at the intersection of hybrid maintenance triggering, uncertain future usage, and full-horizon schedule evaluation under probabilistic consequences.

The objective of this paper is deliberately focused. We do not propose a new large-scale optimization algorithm, nor do we claim a fully realistic industrial maintenance-planning model. Instead, we formulate a unified theoretical framework for predictive maintenance scheduling under uncertainty, intended as a basis for later comparative and computational work. The key idea is to evaluate a complete maintenance schedule through scenario-based realizations of future usage and degradation and to assess that schedule using a cost or utility functional defined on the resulting distribution of outcomes.

The main contributions of the paper are as follows:
\begin{enumerate}
    \item we identify a research gap at the intersection of calendar-based, usage-based, and prognostics-based maintenance decision-making;
    \item we formulate a unified finite-horizon scenario-based framework for maintenance scheduling under uncertainty;
    \item we provide an illustrative computational example showing that integrated scenario-based policies can substantially outperform simpler single-trigger rules in a synthetic setting.
\end{enumerate}

The contribution of the paper should therefore be interpreted primarily as conceptual and methodological, with an accompanying proof-of-concept numerical illustration. In the present example, the main empirical effect is the benefit of integrating heterogeneous maintenance signals within a common scheduling framework, whereas the difference between risk-neutral and risk-aware integrated policies remains comparatively modest.

The remainder of the paper is organized as follows. Section II reviews the relevant literature. Section III introduces the problem setting and assumptions. Section IV defines the compared decision policies and their evaluation criteria. Section V reports the illustrative computational example and its results. Section VI discusses limitations and possible extensions. Section VII concludes the paper.
\section{Related Work}

Maintenance optimization has traditionally been developed along two main lines: time- or age-based maintenance and condition-based maintenance. Early review papers document this distinction and show how maintenance policies evolved from fixed service intervals toward decisions supported by observed condition and degradation information \cite{Ahmad2012}. Broader reviews further show that modern maintenance optimization spans multiple policy classes, information structures, and system scales, including both single-unit and multi-component settings \cite{deJonge2020}. Recent surveys also emphasize that predictive maintenance should be viewed not only as a prognostics task, but as a decision-making problem linking monitoring outputs with operational and maintenance actions \cite{Bousdekis2019,Li2020}.

A substantial part of the predictive maintenance literature focuses on prognostics and remaining useful life estimation. In this stream, RUL is treated as a key indicator supporting intervention timing. Representative works already use RUL predictions directly in maintenance planning, for example in asset-intensive applications such as wind energy systems \cite{Lei2018}. However, many such formulations remain centered on condition or RUL as the dominant maintenance trigger, rather than on joint scheduling based on several heterogeneous signals.

A second important line of work concerns maintenance planning for systems with multiple components or multiple assets. In such settings, maintenance decisions interact through uncertainty, shared interventions, and economic or structural dependencies \cite{OldeKeizer2017}. Finite-horizon stochastic formulations have already been proposed for multi-component maintenance planning under uncertain degradation and uncertain future system evolution \cite{Zhu2021,IslamVatn2023}. Industrial planning studies likewise show growing interest in combining predictive and preventive logic within a common scheduling framework \cite{Einabadi2023}.

Risk-aware formulations provide another relevant perspective. Beyond standard expected-cost optimization, several authors have considered maintenance decisions under utility-based or tail-risk-sensitive criteria, thereby linking predictive maintenance scheduling with stochastic programming and distribution-aware optimization \cite{PedersenVatn2022,BirgeLouveaux2011,Rockafellar2000}. More recent prognostics-oriented studies also show that the credibility of RUL estimates can be incorporated directly into maintenance decision logic \cite{Shi2025}. In parallel, hybrid age-based and condition-based strategies have already been studied for heterogeneous multi-component systems \cite{Bautista2025}.

Despite these developments, the literature remains fragmented. Existing works typically emphasize one dominant maintenance trigger, a specific pairwise hybridization, or a particular optimization setting. What is still less standardized is a unified finite-horizon formulation in which a complete schedule is determined jointly by three sources of information: a calendar-based maintenance limit, a usage-based limit driven by uncertain future cycles, and a condition-monitoring signal represented by uncertain RUL. The present paper is positioned at this intersection and proposes a scenario-based framework for evaluating such schedules under both expected-cost and tail-risk criteria.

\section{Problem Formulation}

\subsection{System, horizon, and decision variables}

Consider a set of assets indexed by $i \in \mathcal{I} = \{1,\dots,N\}$ over a discrete planning horizon $\mathcal{T} = \{1,\dots,T\}.$

The decision problem is to determine a maintenance schedule over the horizon. Let
$
x_{i,t} \in \{0,1\}
$
denote a binary variable equal to 1 if a maintenance action is scheduled for asset $i$ at time instant $t$, and 0 otherwise.

For the simplified setting considered in the present paper, at most one maintenance action is allowed for each asset within the planning horizon:
\[
\sum_{t \in \mathcal{T}} x_{i,t} \leq 1, \qquad \forall i \in \mathcal{I}.
\]

Additional constraints, such as limited crews, maintenance windows, or grouped interventions, are intentionally omitted at this stage and treated as possible extensions.

\subsection{Asset state descriptors}

For each asset $i$, three maintenance-relevant descriptors are considered.
\begin{itemize}
    \item $a_{i,t}$, denotes the calendar age, i.e., the elapsed time since the last maintenance action.
    \item $u_{i,t}$, denotes the cumulative usage, expressed for instance in operating cycles. The future evolution of usage is uncertain.
    \item $r_{i,t}$, denotes the estimated remaining useful life (RUL). In the present paper, $r_{i,t}$ is treated as an uncertain quantity represented through a location-scale description, e.g., by a mean estimate and standard deviation.
\end{itemize}

For each asset, two maintenance resource limits are assumed known $
\bar{a}_i$ and $\bar{u}_i$,
where $\bar{a}_i$ is the recommended calendar-based overhaul interval and $\bar{u}_i$ is the recommended usage-based limit.

Whenever maintenance is performed, the state descriptors are reset. In the simplest form,
\[
a_{i,t+1} =
\begin{cases}
0, & \text{if } x_{i,t}=1,\\
a_{i,t}+1, & \text{otherwise},
\end{cases}
\]
and similarly for cumulative usage,
\[
u_{i,t+1} =
\begin{cases}
0, & \text{if } x_{i,t}=1,\\
u_{i,t} + \Delta u_{i,t}(\omega), & \text{otherwise},
\end{cases}
\]
where $\Delta u_{i,t}(\omega)$ is the uncertain usage increment under scenario $\omega$.

\subsection{Scenario-based uncertainty description}

Let $\omega \in \Omega$ denote a scenario. Each scenario represents a sampled realization of uncertain future evolution, including:
\begin{itemize}
    \item usage increments $\Delta u_{i,t}(\omega)$,
    \item RUL trajectories or sampled latent degradation states,
    \item possibly other quantities affecting maintenance value, such as performance deterioration or failure loss.
\end{itemize}

In this initial formulation, usage processes are assumed independent across assets. This assumption is made for clarity and tractability, and it can later be relaxed by introducing correlated scenario generation.

Let $p_\omega$ denote the probability assigned to scenario $\omega$, with
\[
p_\omega \ge 0, \qquad \sum_{\omega \in \Omega} p_\omega = 1.
\]

\subsection{Failure risk model}

We assume that the failure probability increases as the effective RUL decreases. Let $
P^{\mathrm{fail}}_{i,t}(\omega; x)
$
denote the probability of failure of asset $i$ at time $t$ under scenario $\omega$ and schedule $x$.

At this stage, the exact mapping is not fixed; however, the intended structure is a monotone saturating function of the effective RUL margin. A generic form is
\[
P^{\mathrm{fail}}_{i,t}(\omega; x)
=
\min\left\{0.95,\; \phi_i\!\left(r^{\mathrm{eff}}_{i,t}(\omega; x)\right)\right\},
\]
where $r^{\mathrm{eff}}_{i,t}(\omega; x)$ is the effective remaining useful life after accounting for scheduled maintenance decisions, and $\phi_i(\cdot)$ is a decreasing function of RUL margin, chosen so that failure risk rises as the asset approaches end-of-life.

For example, one may later consider exponential, logistic, or piecewise-linear specifications. For the purposes of this theoretical paper, only the qualitative properties are required:
\begin{enumerate}
    \item $P^{\mathrm{fail}}_{i,t}$ increases as effective RUL decreases,
    \item $P^{\mathrm{fail}}_{i,t}$ remains bounded by $0.95$,
    \item crossing the nominal RUL threshold corresponds to entering a high-risk region rather than to deterministic failure.
\end{enumerate}

\subsection{Cost and utility structure}

For each asset $i$, time $t$, and scenario $\omega$, define the scenario-wise contribution
\[
J_{i,t}(x,\omega)
=
J^{\mathrm{pm}}_{i,t}(x)
+
J^{\mathrm{fail}}_{i,t}(x,\omega)
+
J^{\mathrm{perf}}_{i,t}(x,\omega)
+
J^{\mathrm{early}}_{i,t}(x,\omega),
\]
where:
\begin{itemize}
    \item $J^{\mathrm{pm}}_{i,t}(x)$ is the direct cost of preventive maintenance,
    \item $J^{\mathrm{fail}}_{i,t}(x,\omega)$ is the expected or realized failure-related loss,
    \item $J^{\mathrm{perf}}_{i,t}(x,\omega)$ is a penalty related to performance deterioration,
    \item $J^{\mathrm{early}}_{i,t}(x,\omega)$ penalizes excessively early intervention.
\end{itemize}

A simple example of the failure-related term is
\[
J^{\mathrm{fail}}_{i,t}(x,\omega)
=
C^{\mathrm{fail}}_i \, P^{\mathrm{fail}}_{i,t}(\omega; x),
\]
where $C^{\mathrm{fail}}_i$ denotes the asset-specific failure consequence coefficient.

The performance-related term can be linked to the estimated RUL through a degradation penalty, for example
\[
J^{\mathrm{perf}}_{i,t}(x,\omega)
=
\psi_i\!\left(r^{\mathrm{eff}}_{i,t}(\omega; x)\right),
\]
where $\psi_i(\cdot)$ is chosen to increase as the remaining life becomes small. This allows one to represent not only catastrophic failure risk, but also gradual loss of efficiency or quality.

The total scenario-wise cost is then
\[
J(x,\omega)
=
\sum_{t \in \mathcal{T}} \sum_{i \in \mathcal{I}} J_{i,t}(x,\omega).
\]

\section{Decision Policies and Evaluation Criteria}

To connect the general formulation with the illustrative computational example, we distinguish between baseline single-trigger policies and integrated scenario-based policies. In the present paper, the computational task is limited to selecting one maintenance date per asset within a finite planning horizon.

\subsection{Single-trigger baseline policies}

The first group of policies uses only one maintenance signal at a time.

\textit{Calendar-only policy:} maintenance is scheduled when the elapsed time since the last intervention approaches the calendar-based maintenance limit.

\textit{Usage-only policy:} maintenance is scheduled when the predicted cumulative usage approaches the usage-based maintenance limit.

\textit{RUL-threshold policy:} maintenance is scheduled when the condition-monitoring signal indicates that the remaining useful life is below a chosen threshold or confidence margin.

These three policies serve as simple baselines. They are intentionally myopic in the sense that each of them relies on one dominant maintenance trigger only.

\subsection{Integrated scenario-based policies}

The second group of policies combines all three information sources: calendar age, cumulative usage, and uncertain RUL.

For a candidate schedule $x$ and scenario $\omega$, let $J(x,\omega)$ denote the corresponding scenario-wise total cost. The integrated policies evaluate candidate schedules under a distribution of future outcomes rather than through a single deterministic forecast.

In the illustrative example, the integrated policy is implemented in a simplified form in which one maintenance date is selected for each asset over the horizon. This yields a complete multi-asset schedule, while keeping the computational example transparent.

\subsection{Schedule evaluation criteria}

Two evaluation criteria are considered in the present paper.

\textit{Expected-cost criterion:}
\[
\min_{x \in \mathcal{X}} \mathbb{E}[J(x,\omega)]
=
\min_{x \in \mathcal{X}} \sum_{\omega \in \Omega} p_\omega J(x,\omega).
\]

This criterion is risk-neutral and favors schedules with the smallest average total cost.

\textit{CVaR-based criterion:}
\[
\min_{x \in \mathcal{X}} \mathrm{CVaR}_{\alpha}(J(x,\omega)).
\]

For a cost random variable $Z = J(x,\omega)$, the conditional value-at-risk at level $\alpha \in (0,1)$ is understood here as the expected cost in the upper tail of the distribution beyond the corresponding value-at-risk threshold:
\[
\mathrm{CVaR}_{\alpha}(Z)
=
\mathbb{E}\!\left[ Z \mid Z \geq \mathrm{VaR}_{\alpha}(Z) \right],
\]
where $\mathrm{VaR}_{\alpha}(Z)$ denotes the $\alpha$-quantile of $Z$ distribution.

Thus, minimizing $\mathrm{CVaR}_{\alpha}$ means selecting schedules that reduce the average cost in unfavorable high-cost scenarios rather than only the average cost over all scenarios.

More generally, the framework can also accommodate other mean-risk or utility-based criteria, but these are not explored computationally in the present paper.\subsection{Interpretation}

The key distinction in the computational study is therefore not only between different objective functions, but also between different classes of decision rules. The first class uses a single maintenance trigger, whereas the second class integrates calendar-, usage-, and prognostics-based information within a common scenario-based evaluation framework. As shown in Section V, this distinction is more important in the present example than the difference between the two integrated optimization criteria themselves.

\section{Illustrative Computational Example and Results}

% Cell 14: LaTeX table with results

\begin{table*}[t]
\caption{Summary of policy performance in the illustrative example.}
\label{tab:policy_summary}
\centering
\begin{tabular}{lrrrr}
\toprule
Policy & Expected cost & CVaR$_{0.9}$ & Mean maint. time & Mean failure proxy \\
\midrule
Integrated expected cost & 148.81 & 151.40 & 2.0 & 0.012 \\
Integrated CVaR          & 149.16 & 150.85 & 1.8 & 0.003 \\
Calendar only            & 916.76 & 1499.16 & 6.4 & 6.386 \\
RUL threshold            & 1132.37 & 1831.12 & 8.6 & 7.703 \\
Usage only               & 1296.44 & 1940.00 & 8.0 & 9.420 \\
\bottomrule
\end{tabular}
\end{table*}
\begin{figure}[t]
    \centering
    \includegraphics[width=\linewidth]{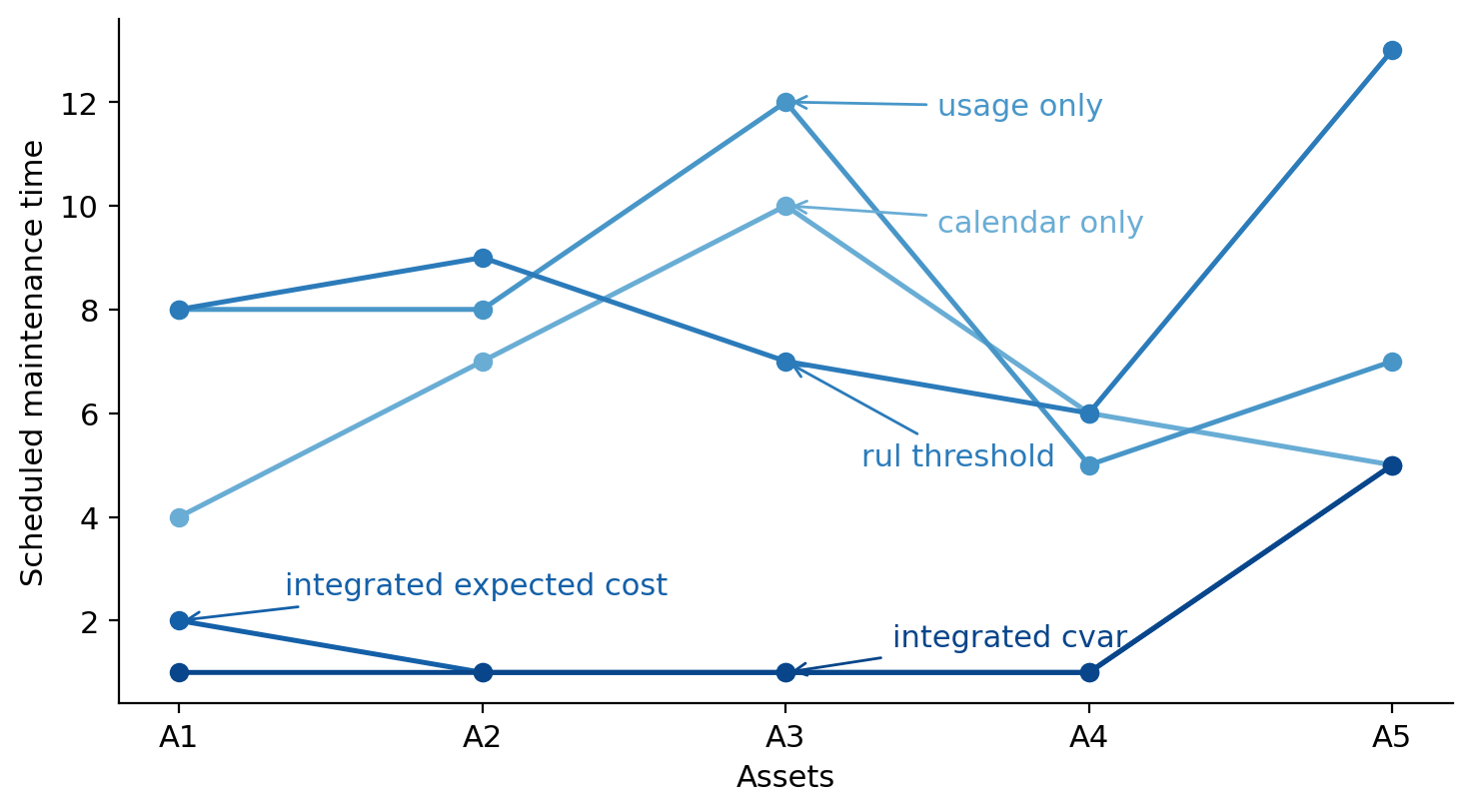}
    \caption{Scheduled maintenance times for the five compared policies. The integrated policies select substantially earlier interventions than the single-trigger rules for most assets.}
    \label{fig:timing}
\end{figure}

\begin{figure}[t]
    \centering
    \includegraphics[width=\linewidth]{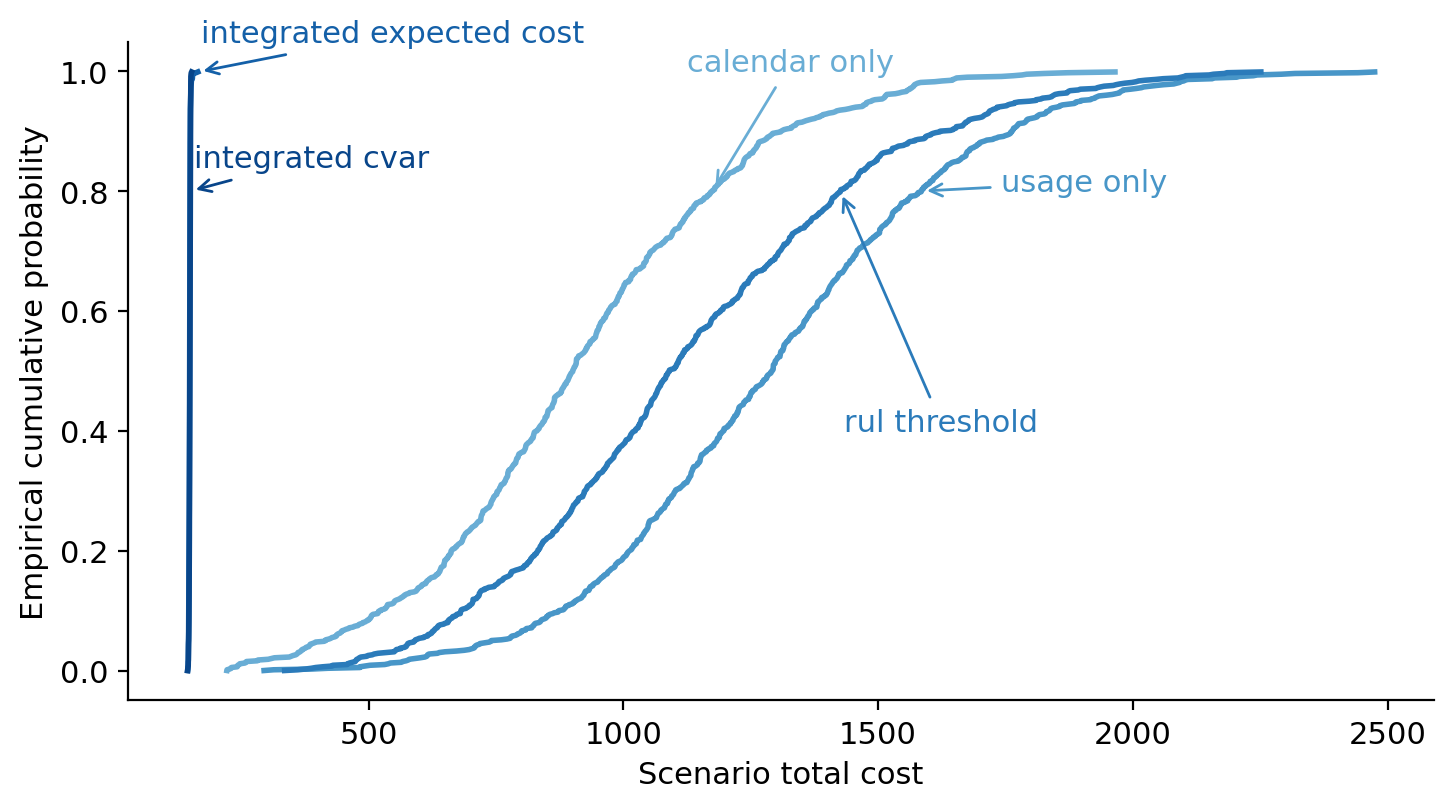}
    \caption{Empirical cumulative distributions of total scenario cost. The integrated policies are strongly shifted to the left and exhibit much narrower cost distributions than the single-trigger rules.}
    \label{fig:ecdf}
\end{figure}

To make the proposed framework operational, we constructed a small synthetic example involving five assets over a finite planning horizon. The purpose of this example is not to provide a realistic industrial benchmark, but to verify whether the proposed scenario-based formulation yields different scheduling decisions than simpler single-trigger policies.

The illustrative experiment considered $N=5$ assets over a horizon of $T=12$ discrete periods and used $800$ Monte Carlo scenarios. Asset parameters were generated synthetically from predefined intervals to obtain a heterogeneous small fleet: calendar limits were sampled from $[8,16]$, usage limits from $[160,320]$, mean RUL values from $[4,13]$, and RUL standard deviations from $[0.8,2.4]$. Initial ages and cumulative usage levels were initialized as random fractions of the corresponding limits. Future usage increments were generated independently across assets and scenarios from positive-valued distributions with asset-specific mean usage in $[10,22]$ cycles per period and coefficient of variation in $[0.15,0.35]$, while latent scenario RUL values were sampled from truncated normal distributions. The failure-risk proxy used a saturating model with maximum level $0.95$, decay rate $0.75$, and performance window equal to $4$ periods. The RUL-threshold baseline used a trigger probability of $0.60$, and the integrated risk-aware policy minimized $\mathrm{CVaR}_{0.9}$.

Each asset was characterized by a calendar-based maintenance limit, a usage-based limit expressed in cumulative cycles, and an uncertain RUL estimate summarized by a mean and standard deviation. The current age and accumulated usage were initialized as fractions of the corresponding asset-specific limits. Future usage increments were then generated scenario-wise from positive-valued distributions calibrated by asset-specific mean usage and variability parameters, while the latent scenario RUL values were sampled from truncated normal distributions. The resulting scenario set was used to evaluate the cost of candidate maintenance dates under the cost model.

Five decision rules were compared: calendar-only scheduling, usage-only scheduling, RUL-threshold scheduling, integrated expected-cost scheduling, and integrated CVaR-based scheduling. The first three act as single-trigger baselines. The last two use the full scenario-based cost representation and differ only in the optimization criterion: expectation versus tail-risk minimization.

The resulting maintenance times are shown in Fig.~\ref{fig:timing}. The clearest pattern is that the two integrated policies schedule maintenance substantially earlier than the single-trigger rules for most assets. In fact, the integrated expected-cost and integrated CVaR policies differ only for one asset: the latter advances maintenance for Asset A1 from period 2 to period 1, while the remaining maintenance dates are identical. This indicates that, under the present calibration, the dominant effect is the integration of heterogeneous maintenance information rather than a strong separation between risk-neutral and risk-aware integrated objectives.

The quantitative comparison is summarized in Table~\ref{tab:policy_summary}. The integrated expected-cost policy achieved the lowest expected total cost, equal to 148.81, while the integrated CVaR policy produced a very similar expected cost of 149.16. At the same time, the CVaR-based policy achieved a slightly lower tail-risk measure, with $\mathrm{CVaR}_{0.9}=150.85$ compared to 151.40 for the expected-cost formulation. Both integrated policies also yielded very small mean failure proxies and substantially earlier average maintenance times than the single-trigger rules.

By contrast, all three single-trigger policies were markedly more expensive and substantially more risk-exposed. Among them, the calendar-only rule performed best, but its expected total cost still reached 916.76 and its $\mathrm{CVaR}_{0.9}$ reached 1499.16. The RUL-threshold and usage-only rules were even less favorable, with expected costs above 1100 and much heavier right tails.

These differences are visible directly in Fig.~\ref{fig:ecdf}, which reports the empirical cumulative distributions of total scenario cost. The two integrated policies produce narrow distributions concentrated at low cost levels, whereas the single-trigger policies generate broad distributions shifted far to the right. This indicates that combining calendar-, usage-, and RUL-based information within a common scenario-based objective reduces average cost and the dispersion of outcomes.

Taken together, the results support two observations. First, even a minimal integrated scenario-based formulation can strongly outperform maintenance rules based on a single trigger. Second, in the present synthetic example, the difference between minimizing expected cost and minimizing tail risk is real but modest. Accordingly, the illustrative study highlights the value of information integration more strongly than the value of risk aversion itself.

% Cell 17: final Section VI

\section{Discussion and Limitations}

The illustrative example confirms that the proposed framework is operational and that it produces schedules clearly different from those implied by calendar-only, usage-only, or RUL-threshold rules. This is the main positive result of the computational study. Even in a small synthetic setting, the integrated scenario-based policies lead to substantially lower expected cost, substantially lower tail risk, and much tighter empirical cost distributions.

At the same time, the numerical example should be interpreted carefully. The dominant finding is not a strong divergence between risk-neutral and risk-aware integrated optimization. Rather, the principal gain comes from combining heterogeneous maintenance signals within a common decision rule. The integrated expected-cost and integrated CVaR schedules are almost identical, and their performance differs only marginally. This suggests that, under the present calibration, the main driver of improvement is the avoidance of clearly unfavorable delayed-maintenance states, whereas tail-risk aversion plays a secondary role.

A more fundamental limitation concerns the current cost structure. In the present implementation, scenario-wise costs are accumulated up to the chosen maintenance date, while the post-maintenance evolution of the asset is not modeled explicitly within the same horizon. As a result, the formulation naturally tends to favor relatively early interventions whenever the penalty for delaying maintenance dominates the penalty for acting too early. This mechanism helps explain why the integrated policies place most maintenance actions in the first few periods. The current example should therefore be interpreted primarily as a proof of concept, not as a quantitatively realistic maintenance-planning benchmark.

The synthetic data-generating process is another simplification. Future usage is generated independently across assets, and RUL uncertainty is represented in reduced form rather than derived from a fully specified probabilistic degradation model fitted to observations. This keeps the example transparent and light, but it also limits the realism of the uncertainty structure. In particular, the current setup does not represent correlated operating regimes, sequential Bayesian updating of prognostic beliefs, or richer condition-monitoring trajectories.

The present formulation also omits several practically relevant scheduling features, such as shared maintenance crews, maintenance windows, grouped shutdowns, or repeated maintenance opportunities within the same horizon. These omissions are deliberate. Their inclusion would make the optimization problem considerably richer, but would also obscure the central purpose of the present paper, which is to define and illustrate a unified scenario-based decision framework.

Despite these limitations, the current example is sufficient for an initiating conference paper. It demonstrates that the proposed framework can produce interpretable maintenance schedules and that these schedules differ materially from single-trigger baselines. It also shows that the framework can accommodate both expected-cost and tail-risk criteria, even if the difference between them remains limited in the present calibration. The most natural next steps are therefore clear: refine the cost model to represent post-maintenance evolution explicitly, replace reduced-form uncertainty generation with richer probabilistic models, and extend the optimization problem toward resource-coupled multi-asset scheduling.

% Cell 20: replacement for Section VII (Conclusions)

\section{Conclusions}

This paper addressed predictive maintenance scheduling under uncertainty in a setting where three heterogeneous maintenance signals are available simultaneously: calendar-based maintenance limits, usage-based limits driven by uncertain future operating cycles, and uncertain condition-monitoring information represented through RUL estimates. The main objective was not to develop a fully realistic industrial scheduling model or a new large-scale optimization algorithm, but to formulate a coherent finite-horizon decision framework capable of integrating these signals within a common scenario-based perspective.

The proposed formulation treats maintenance planning as a schedule-selection problem evaluated under a distribution of future outcomes rather than under a single deterministic forecast. This makes it possible to compare both expected-cost and risk-aware policies within the same modeling structure. In this sense, the main contribution of the paper is conceptual: it clarifies how calendar-, usage-, and prognostics-based information can be brought together in a unified scheduling framework.

The illustrative computational example supports the practical relevance of this perspective. Even in a small synthetic setting, the integrated scenario-based policies substantially outperformed simpler calendar-only, usage-only, and RUL-threshold rules. The main empirical gain came from integrating heterogeneous maintenance information in a common decision rule. By contrast, the difference between the expected-cost and CVaR-based integrated policies remained limited under the present calibration, suggesting that the value of risk-aware optimization is more subtle in this particular proof-of-concept setting than the value of information integration itself.

The present results should be interpreted as a proof of concept rather than as a realistic maintenance benchmark. Nevertheless, they show that integrating disparate information within a common scenario-based framework is a promising direction for predictive maintenance. Future work should focus on richer cost dynamics, probabilistic degradation models fitted to data, and resource-coupled multi-asset scheduling.

\balance
\bibliographystyle{IEEEtran}
\bibliography{references}
\end{document}